\def\rect#1#2{{\vcenter{\vbox{\hrule height.3pt
	    \hbox{\vrule width.3pt height#2truecm \kern#1truecm
	    \vrule width.3pt}
	    \hrule height.3pt}}}}

\def\aversim#1#2{\lower3pt\vbox{\baselineskip0pt \lineskip-.1pt
    \ialign{$\mathsurround=0pt #1\hfil##\hfil$\crcr#2\crcr\sim\crcr}}}

\documentstyle[12pt]{article}
\begin{document}
\begin{center}
\LARGE
Folding transition of the triangular lattice in a discrete
three--dimensional space
 ~\\
 ~\\
 \normalsize
 ~\\
\normalsize
Emilio N.M. Cirillo\\
{\it Dipartimento di Fisica dell'Universit\`{a} di Bari} and\\
{\it Istituto Nazionale di Fisica Nucleare, Sezione di Bari\\
via Amendola 173, 70126 Bari, Italy}
 ~\\
 ~\\
\normalsize
Giuseppe Gonnella$^{(*)}$\\
{\it Theoretical Physics, Oxford University, 1 Keble Rd Oxford OX1 3NP, UK}
{}~\\
{}~\\
\normalsize
Alessandro Pelizzola\\
{\it Dipartimento di Fisica del Politecnico di Torino} and \\
{\it Istituto Nazionale per la Fisica della Materia, \\
c. Duca degli Abruzzi 24, 10129 Torino, Italy}
 ~\\
 ~\\
\end{center}
\vskip 2.0cm
\begin{abstract}
A vertex model introduced by M. Bowick, P. Di Francesco, O. Golinelli,
and E. Guitter [Nucl. Phys. B {\bf 450}, 463 (1995)]  describing
the  folding of the triangular
lattice onto the  face centered cubic lattice
has been studied in the hexagon
approximation of the cluster variation method.
The model  describes the behaviour of a polymerized membrane in a discrete
three--dimensional space.
We have introduced a curvature energy and a symmetry breaking field
and  studied the phase diagram of the resulting model.
By varying the curvature energy parameter,  a first-order
transition has been found between a flat and a folded phase for  any value
of the symmetry breaking field.
\end{abstract}
\vskip  1.5cm
\noindent
PACS numbers:
05.50.+q (Ising problems);
64.60.-i (General studies of phase transitions);
82.65.Dp (Thermodynamics of surfaces and interfaces).
\vskip 1.5 cm
$^{(*)}$ Permanent Address:\\
{\it Dipartimento di Fisica dell'Universit\`{a} di Bari} and\\
{\it Istituto Nazionale di Fisica Nucleare, Sezione di Bari,
via Amendola 173, 70126 Bari, Italy}

\newpage

\addtolength{\baselineskip}{\baselineskip}

The macroscopic behaviour of membranes fluctuating in the euclidean space
crucially depends on the microscopic characteristics of the system
\cite{NPW}.
For example, a {\it fluid} membrane without self-avoiding interactions is
always expected to be in a crumpled state \cite{PL}, independently
from the stiffness attributed to the membrane,
while a rigid phantom \cite{pli} {\it polymerized}
 membrane, which is a network of
molecules with fixed connectivity, is expected to be stable in
a flat phase \cite{NP}.
The existence of different classes of membrane systems
suggests to consider specific lattice models paradigmatic
for each class, which can be useful for analytical and numerical
calculations.
While many--component membrane systems can be sometimes expressed
as Ising--like models \cite{ccgm}, it is more difficult to describe
the statistical behaviour of a single membrane in terms of
usual lattice models with a local hamiltonian.
In Ref. \cite{npb95}, a $D=2$ vertex model has been introduced which
describes the behaviour of a single phantom  polymerized membrane
with bonds of fixed length embedded in a discrete $d=3$ space.
The aim of this communication is to study the phase diagram of the
membrane model of \cite{npb95}.

Models of polymerized membranes were introduced
in \cite{KN}, and studied using a Landau--Ginzburg evaluation
of the energy of the system in \cite{NP,PKN,DG,AL}.
{}From the above studies,
and from numerical simulations \cite{BEW}, it comes out that,
when excluded volume effects are not considered,
 by varying the strength $K$ of a bending
energy term which favours flat configurations, a critical transition
arises separating a flat phase at large $K$ from a crumpled
phase at small $K$.

Generally, the length of the bonds of a polymerized membrane can
vary  being subjected to elastic forces \cite {NP}.
In \cite{KJ} the simple case of a triangular network with bonds of
fixed length embedded in a $d$--dimensional space
has been first considered. This folding problem has been studied
in \cite{epl,pre,nostro}
in the case of a two--dimensional embedding space; here the normals
to the triangles of the network can point only ``up'' or ``down''
in some direction. This model can be mapped on a $11$--vertex model
equivalent to a constrained Ising model with some
spin configurations forbidden \cite{epl};  a first--order transition
has been found to occur between a flat and a disordered phase
\cite{pre,nostro}.

The more complicated problem of
the folding of the triangular lattice in a
three--dimensional embedding space has been
formulated in \cite {npb95}, with the embedding space
discretized as a face centered cubic lattice.
 In this model the
plaquettes of the triangular lattice are mapped, by folding, onto those
of a face centered cubic lattice, so that two adjacent
plaquettes form an angle which can take up only four different values
(see Fig. 1).
%The folded lattice is assumed to be a phantom one, i.e.
%self--intersections are allowed.
%, since the self--avoidance constraint
%would lead to highly non--local interactions which cannot be taken into
%account in the present framework.
In the following we introduce a term representing the stiffness
of the network and study the phase diagram of the resulting model
by applying the cluster variation method (CVM)
in the same approximation used in \cite{nostro}.

The vertex model of \cite{npb95} can be defined as follows.
Two ${\bf Z}_2$ variables, named $z_i$ and $\sigma_i$,
are   assigned  to each plaquette of the
triangular lattice, and hence to each site $i$ of the dual hexagonal
lattice.
The relative values of these variables on adjacent
plaquettes (say 1 and 2) determine the angle that is formed by the
plaquettes, according to the following rules: for $z_1 = z_2$ and
$\sigma_1 = \sigma_2$ we have no fold between the plaquettes; for $z_2 =
- z_1$ and $\sigma_2 = \sigma_1$ we have an acute fold, with an angle
$\theta$ such that $\cos \theta = 1/3$; for $z_2 = - z_1$ and $\sigma_2
= - \sigma_1$ we have an obtuse fold, with $\cos \theta = - 1/3$;
finally, for $z_2 = z_1$ and $\sigma_2 = - \sigma_1$ we have a complete
fold, with the plaquettes lying one on top of the other (see Fig. 1).

It has been shown in \cite{npb95} that the variables $z_i$ and
$\sigma_i$ have to satisfy two constraints, or folding rules, in order
to describe a folding configuration over the face centered cubic
lattice. Such folding rules take the form
\begin{equation} \sum_{i=1}^6
\sigma_i = 0 \ {\rm mod}\  3, \label{rule1}
\end{equation}
where the index $i$ runs around a hexagon, and
\begin{equation}
\sum_{i=1}^6 \frac{1 - z_i z_{i+1}}{2} \ \Delta_{i,c}
= 0 \ {\rm mod}\ 2 \;\;\;\; c=1,2 \quad ,
\label{rule2}
\end{equation}
where $z_7=z_1$ and
\begin{equation}
\Delta_{i,c}=\left\{
\begin{array}{ll}
1& \;\;\; {\rm if}\; \sum_{j=1}^i \sigma_j =  c\  {\rm mod}\ 3 \\
0& \;\;\; {\rm otherwise}
\end{array}
\;\;  i = 1,...,6,\; c = 1,2
\right.
\;\; .
\end{equation}
\par
We shall assume that a fold between two adjacent plaquettes has an
energy cost, due to curvature, given by $-K \cos \theta$, where $\theta$ is the
angle
between the normal vectors to the plaquettes. In terms of our
Ising--like variables such a term can be rewritten as $-K \sigma_i
\sigma_j (1 + 2 z_i z_j)/3$, manifestly symmetric under the global
transformations $z_i \rightarrow - z_i, \forall i$ and $\sigma_i
\rightarrow -\sigma_i, \forall i$. We will introduce
 also a term which breaks this symmetries, the analog of the
magnetic field in the ordinary Ising model. In the two--dimensional case
\cite{pre,nostro} it was quite easy to define a symmetry--breaking field
coupled
to the direction of the normal to a plaquette, since there the
spin variable associated to each plaquette denoted whether the normal
was pointing up or down. In the present case and with the present
parametrization this is no more possible, since the pair of values
$\{z_i, \sigma_i\}$ does not determine the orientation of the plaquette.
However, we are not interested in a true magnetic field, which cannot be
given a physical meaning in this model, but just in a symmetry--breaking
term, which can be realized in several ways. A simple choice, which
favors only one of the four possible flat states (a flat state is easily
seen to be characterized by the condition $\{z_i, \sigma_i \}$
independent of $i$), leaving on the same ground the remaining three, is
$-h \delta_{z_i,1} \delta_{\sigma_i,1}$. There are of course other
possibilities, and in the following we shall consider one of these in
order to show that the phase diagram is, roughly speaking, qualitatively
independent of this choice.

We are thus led to consider the
following hamiltonian (energies are given in units of $k_{\rm B}T$),
defined on the dual hexagonal lattice:
\begin{equation}
{\cal H} = -\frac{K}{3} \sum_{\langle i j \rangle} \sigma_i \sigma_j (1
+ 2 z_i z_j) - h \sum_i \delta_{z_i,1} \delta_{\sigma_i,1},
\label{hamilton}
\end{equation}
where the first sum is over nearest neighbor pairs.

As for the two--dimensional model, the phase diagram will be
investigated by means of the hexagon approximation of the cluster
variation method (CVM), which has been thoroughly described in
\cite{nostro}. This requires to introduce a hexagon density matrix
$\rho_6(\{z_i, \sigma_i\})$, indexed by the hexagon configurations, a
link density matrix $\rho_2$ and two site density matrices $\rho_{1A}$
and $\rho_{1B}$, one for each of the two interpenetrating sublattices of
the hexagonal lattice.
Because of the folding rules Eqs.\ \ref{rule1} and \ref{rule2} one has
not to consider $2^{12} = 4096$ hexagon configurations, but only 384,
and hence 384 $\rho_6$ elements (this
number may be slightly reduced by taking into account degeneracies, but
this would lead to a more involved formulation).
The link and site density matrices can be defined as partial traces of
$\rho_6$ and, in analogy with \cite{nostro}, we can
write the variational free energy density (in units of $k_{\rm B} T$)
\begin{eqnarray}
f &=& - \frac{K}{2} {\rm Tr}[(\sigma_1 \sigma_2 (1 + 2 z_1 z_2) \rho_2]
- \frac{h}{2} [\rho_{1A}(z=1,\sigma=1) + \rho_{1B}(z=1,\sigma=1)]\nonumber \\
&& + \frac{1}{2} {\rm Tr}(\rho_6 \ln \rho_6) - \frac{3}{2} {\rm Tr}
(\rho_2 \ln \rho_2) + \frac{1}{2} \left[ {\rm Tr} (\rho_{1A} \ln
\rho_{1A}) + {\rm Tr} (\rho_{1B} \ln \rho_{1B}) \right]  \\
&& + \lambda({\rm Tr}\ \rho_6 - 1),\nonumber
\end{eqnarray}
where $\lambda$ is a Lagrange multiplier which ensures the normalization
of $\rho_6$ (and hence also of $\rho_2$ and $\rho_{1A}, \rho_{1B}$).
This variational functional must, in general, be minimized numerically,
and this can be done easily by an iterative method, as explained in
\cite{nostro}.

At infinite temperature (or vanishing $K$ and
$h$) we obtain the entropy (per site) $S = \ln q$, where $q
\simeq 1.42805$, in very good agreement with the transfer matrix estimate $q =
1.43(1)$ found in \cite{npb95}, although slightly lower than the best
lower bound 1.43518 obtained from the analysis of two--dimensional
folding in a staggered field \cite{npb95}. Furthermore we obtain
$\langle z_i \rangle_A = \langle z_i \rangle_B = 0$ and
$\langle \sigma_i \rangle_A = -\langle \sigma_i \rangle_B \simeq
0.87456$, which, together with the values of the link density matrix
elements, indicate a marked preference of the triangular lattice for
obtuse and complete folds.

Let us now describe the main features of the phase diagram of the model
Eq.\ \ref{hamilton}, which has been obtained by means of the CVM and
reported in Fig. 2, in the $K-h$ plane.
The solid line separates the flat phase (high values of $K$ and $h$, or
low temperature) from the
disordered, or folded, one (low values of $K$ and $h$, or high temperature).
The transition is of first
order on the whole line; on the $h=0$ axis the transition occurs at
$K=0.18548$, while on the $K=0$ axis it occurs at $h=0.49839$. The two phases
can be distinguished by means of the energy--like correlation function
$\langle \sigma_i \sigma_j (1 + 2 z_i z_j)/3 \rangle$, which is negative in the
folded phase and saturates to 1 in
the flat phase, indicating that the plaquettes are all parallel. As a
consequence, in the flat phase
the entropy vanishes and the free energy reduces to
the internal energy contribution,
\begin{equation}
f_{\rm flat} = - \frac{3}{2} K - h.
\end{equation}
The phase diagram of Fig. 2 is then easily obtained by comparing the
value of the
free energy of the disordered phase with $f_{\rm flat}$ above.

The  first point to be discussed
is that  enlarging the embedding space has not
turned the phase transition into a second order one, in contrast with
what one could expect \cite{npb95}. On the contrary, the
first order character of the $h = 0$ transition seems to be enhanced in
the three--dimensional case, as suggested by the jump in the
energy--like correlation function,
which is about 1.237 in the present case against 1.047 in the
two--dimensional case.
These features should be described correctly by the CVM approximation,
as suggested by the good agreement with the transfer matrix results
obtained both in the two--dimensional case \cite{nostro}
and, for the entropy at infinite temperature, in the present case.

The curvature term in Eq.\ \ref{hamilton}, without the constraints Eqs.\
\ref{rule1}--\ref{rule2}, corresponds to the anisotropic  Ashkin--Teller
model \cite{Bax} with a particular choice of the parameters. We have
studied in the CVM pair approximation the Ashkin--Teller model with the
hamiltonian ${\cal H}_{\rm AT} = -K/3 \sum_{<ij>}\sigma_i \sigma_j (1 + 2 z_i
z_j)$ and we have found two critical transitions at $K_{c1} = 0.824$
(where the symmetries $z \to -z$ and $\sigma \to -\sigma$ are separately
broken, but their product is preserved) and $K_{c2} = 1.648$ (where even
the product symmetry gets broken). In analogy with the result obtained
in the two--dimensional case \cite{nostro}, we see that the introduction
of defects which relax the constraints is necessary to smooth the
transition.

Finally, it has already
been observed before  that the introduction of a
symmetry--breaking field in the hamiltonian is not a trivial task as in the
two--dimensional case \cite{nostro}. It is therefore worth asking
whether the basic features of our phase diagram depend on the choice of
this field. For this reason we have considered a second
model by introducing the $h$--field term in the hamiltonian in the following
way
\begin{equation}
{\cal H}_{\rm alt} = -\frac{K}{3} \sum_{\langle i j \rangle} \sigma_i \sigma_j
(1
+ 2 z_i z_j) - \frac{h}{3} \sum_i \sigma_i(1+2z_i),
\label{hamilton2}
\end{equation}
where $h$ is non--negative. This amounts to let all the
plaquettes interact, with an energy $h$, with a ``ghost'' plaquette
which is fixed in the state characterized by $(z = 1, \sigma = 1)$.
In Fig. 2 it is shown the phase diagram of the model Eq. \ref{hamilton2} as
well; the transition line is the broken one and is still a first order one.
It is clear that the main features
of the phase diagram have not been changed by replacing the old $h$--field
term by the new one.

To summarize, we have studied in the CVM hexagon approximation the phase
diagram of a vertex model which  describes the folding of a triangular
network onto a fcc lattice, subjected to a bending energy and a
symmetry--breaking field. We have determined the
folding entropy at infinite temperature, which is in very good agreement
with a previous transfer matrix estimate, and the phase diagram of the
model, which turns out to be qualitatively independent on the choice of
the symmetry--breaking field. The flat and the folded phases are
separated by a first--order transition, which is even stronger than that
obtained with a two--dimensional embedding space.
The lack of critical behaviour, which is expected for
a polymerized phantom membrane, may be due to the particular
choice of the discrete embedding space.

One of us (G.G.) thanks Amos Maritan for a discussion about the
subject of this paper.

\newpage
\Large
\par\noindent
{\bf Figure Captions}
\normalsize
\addtolength{\baselineskip}{\baselineskip}
\vskip 1 truecm
\par\noindent
Figure 1: The four possible foldings of two adjacent plaquettes of
the triangular lattice embedded in a fcc lattice are shown.
{}From the left to the right and from the top to the bottom:
no fold, acute fold ($70^032'$), obtuse fold ($109^028'$) and complete fold.
The  dots represent
the vertices in an elementary cell of the
fcc lattice.
\vskip 0.5 truecm
\par\noindent
Figure 2: Phase diagram in the $K-h$ plane: the solid and dashed lines
correspond to hamiltonians ${\cal H}$ and ${\cal H}_{\rm alt}$, respectively.


\begin{thebibliography}{99}
\bibitem{pli} Membranes where effects of excluded volume are not
considered are sometimes called phantom membranes.
\end{thebibliography}
\end{document}